\documentclass{crc-book}

\usepackage{float}
\usepackage{amsthm,,amssymb, color, epsfig, graphics}
\usepackage{xcolor}
\usepackage{graphicx}
\usepackage[intlimits]{amsmath}
\usepackage{latexsym}

\theoremstyle{definition}

\begin{document}

%
\renewcommand{\evenhead}{B. Grammaticos et al.}
\renewcommand{\oddhead}{Detecting discrete integrability}

%
\thispagestyle{empty}

\Name{Detecting discrete integrability: the singularity approach}

\Author{B. Grammaticos~$^a$, A. Ramani~$^a$, R. Willox~$^b$ and T. Mase~$^b$}

\Address{$^a$ IMNC, Universit\'e Paris VII \& XI, CNRS, UMR 8165, B\^at. 440, 91406 Orsay, France\\[10pt]
$^b$ Graduate School of Mathematical Sciences, the University of Tokyo, 3-8-1 Komaba, Meguro-ku, 153-8914 Tokyo, Japan}

\begin{abstract}
  \noindent
We describe the various types of singularities that can arise for second order rational mappings and we discuss the historical and present-day, practical, role the singularity confinement property plays as an integrability detector. In particular, we show how singularity analysis can be used to calculate explicitly the dynamical degree for such mappings. \footnote{\bf This work is dedicated to the memory of professor K.M. Tamizhmani, collaborator and beloved friend}
\end{abstract}

\section{Introduction}\label{sec-intro}
The history of integrability, in its broadest sense, goes back to the beginnings of differential calculus in the 17th century and to the quest for solutions to differential equations. Mathematical modelling of physical problems led the founding fathers of calculus, Leibniz, Newton and the Bernouillis, to the study of differential equations and their solutions. The domain blossomed over the next two centuries through the contributions of many great mathematicians like Euler, Lagrange, Gauss, to name but a few. What can be considered as the real beginning of the modern era of integrability though, is the pioneering work of Kovalevskaya on the heavy top, spinning around a fixed point. Kovalevskaya noticed that for all cases where a solution to the equations of motion was known, this solution was given in terms of meromorphic functions and in particular elliptic functions. She was thus led to investigate the existence of other cases with meromorphic solutions and actually discovered one previously unknown case, which was subsequently dubbed the Kovalevski top in her honour.

Meromorphicity of solutions lay also at the origin of Painlev\'e's approach to the construction of new functions through the solution of nonlinear differential equations. Painlev\'e first studied the first-order case showing that the only equation without movable (i.e. initial-condition dependent) critical (i.e. multivaluedness inducing) singularities was the Riccati equation. The Riccati equation, however, can be transformed to a linear equation and therefore does not introduce any new functions, since linear equations are regarded as solvable, essentially, in terms of `known' functions. Painlev\'e therefore went on to study second-order equations, requiring that their solutions be meromorphic in the independent variable apart from possible fixed critical singularities, which can easily be taken care of. This intuition was brilliant since it turned out that several nontrivial such examples do exist and that new functions can be introduced by some of the equations he derived. These new functions are since known as the Painlev\'e transcendents. But even more important than the discovery of the {\em Painlev\'e equations} these functions satisfy, was the realisation that the absence of movable critical singularities (a property which later came to be known as the {\em Painlev\'e property}) could provide a powerful integrability criterion for differential systems \cite{physrep}.

Curiously, the interest in integrable differential systems waned over the next half century, a situation which only changed with the advent of electronic computers and the possibilities for numerical simulation these offered. While studying the famous Fermi-Pasta-Ulam (FPU) model, describing particles interacting on a lattice, Kruskal and Zabusky \cite{kruskal} considered its continuum limit and found that it led to a partial differential equation, introduced at the end of the 19th century by Korteweg and de Vries (known today as the KdV equation), which possesses an explicit  solitary wave solution. Kruskal and Zabusky observed through simulations that the evolution of an initial profile led to its separation into several solitary waves that interact elastically. They chose the name of `soliton' for these special waves. Analytical studies soon followed, showing that the KdV equation possesses infinitely many conservation quantitites and, as was shown by Hirota, allows for an arbitrary number of solitons. The complete solution of KdV was subsequently obtained through the use of methods of quantum mechanical inverse scattering. One important result of these early studies was the observation that Painlev\'e equations often appear as reductions of integrable evolution equations. This led, on the one hand, to the integration of the Painlev\'e equations through inverse scattering methods and, on the other hand, to the formulation of the Ablowitz-Ramani-Segur conjecture \cite{ars}. The ARS conjecture reaffirmed the Painlev\'e property as an integrability criterion, positing that ``every ordinary differential equation which arises as a reduction of a partial differential equation integrable by inverse scattering techniques, is of Painlev\'e type''.

While its continuum limit is integrable, the same is not true for the FPU model itself. An important question therefore was whether one could find a nonlinear lattice that is completely integrable. Toda \cite{toda} showed that this is indeed possible, by introducing a lattice with exponential interactions between nearest neighbours (the system that now bears his name). The Toda lattice is however a semi-discrete system since it describes the positions of particles on a lattice as continuous functions in time. But what about fully discrete systems? In fact, while studying the possible discretisations of the logistic equation (which is a simplified, constant coefficient, Riccati), Skellam \cite{skellam} and Morishita \cite{moris} obtained a discrete form that has the fundamental property of the continuous one: it can be transformed into a linear equation. This discovery, however, remained mainly unnoticed, which was also the case for the groundbreaking work of Hirota who, in the 1970s, singlehandedly produced the integrable, fully discrete, forms of a host of famous integrable partial differential equations \cite{hirota}. 

Fortunately though, this lack of interest was short-lived and at the end of the 80's several findings finally brought about the discrete integrability epoch. One important observation was that Baxter's solution to the Yang-Baxter equations is associated to the Euler-Chasles correspondence \cite{veselov}. This result was cast in integrable systems parlance by Quispel, Roberts and Thompson (QRT) who defined a family of integrable second-degree mappings \cite{qrt} solvable in terms of elliptic functions. Around the same time, work on string theory led to the discovery of integrable non-autonomous recursion relations which turned out (upon derivation of their continuum limits) to be discrete analogues of the Painlev\'e equations \cite{dps}. All this made integrability specialists turn to the until then largely unexplored discrete domain, developing techniques that were, at times, in perfect parallel to those for continuous systems, but on other occasions specific to the discrete case. One important question was ``what is the discrete analogue of the Painlev\'e property'' or, to put it in a less pretentious way, ``is there an easy-to-use discrete integrability criterion''. Our answer to this question, known under the name of {\em singularity confinement property} \cite{sincon} will be the subject of this chapter.

\section{Singularity confinement}\label{sec-singconf}
Let us start with an example of what may happen when one iterates an integrable mapping. For this we shall use the McMillan mapping \cite{mcmillan}, which we shall consider over $\mathbb{P}^1 = \mathbb{C}\cup\{\infty\}$ (and where $\mu$ is an arbitrary non-zero complex number):
\begin{equation}x_{n+1}+x_{n-1}={2\mu x_n\over 1-x_n^2}.\label{zena}\end{equation}
Suppose that, due to a special choice of initial condition, at some iteration step, $x_{n+1}$ takes precisely the value 1, while $x_{n}$ is generic. This leads to values $x_{n+2}=\infty$, $x_{n+3}=-1$ and an indeterminate result for $x_{n+4} : \infty-\infty$. The way to lift this indeterminacy is through continuity with respect to the initial conditions, by introducing a small quantity $\epsilon$ and assuming that $x_{n+1}=1+\epsilon$. If one perfomes the above iteration for this initial condition, taking the limit for $\epsilon\to0$ we then find that $x_{n+4}=-x_{n}$. In other words, by lifting the indeterminacy that arose at step $n+4$, we in fact recovered the memory of the initial condition that was lost at the level of $x_{n+2}$. The loss of a degree of freedom, here the memory of the value of $x_{n}$, constitutes a {\em singularity} for the mapping. Note that this loss of memory of the initial condition in fact means that the inverse mapping is not defined at this point, which is another possible definition of a singularity. 
It should be stressed that a degree of freedom lost in a singularity can only be recovered if, when iterating beyond the singularity, an indeterminacy appears that can be lifted in the way we just described. This is what we call {\em confinement} of the singularity.

The behaviour of the McMillan mapping can, in fact, be easily understood once one realises that (\ref{zena}) is just the addition formula for elliptic sines. Indeed, if we consider a discretisation of the continuous variable $t_n=n\delta+t_0$, we have $x_n=x(t_n)$, $x_{n\pm1}=x(t_n\pm\delta)$ and the solution of (\ref{zena}) is simply $x_n=k\, {\rm sn}\, \delta\,{\rm sn}(t_n)$ (where $k$ is the modulus of the elliptic sine) with $\mu={\rm cn}\delta\, {\rm dn}\delta$. Using the addition formulae we can easily verify that if $x_n$ diverges then $x_{n\mp1}=\pm1$ and $x_{n+2}=-x_{n-2}$. Thus $x_{n-1}$ has precisely the value that guarantees the divergence of $x_n$ and for $x_{n+1}$ the value that compensates this divergence. Moreover, the memory of the value of $x_{n-2}$ survives past the singularity and is recovered at $x_{n+2}$.

The McMillan mapping is not the only one that can be solved in terms of elliptic functions. As pointed out in the introduction, a whole family of such mappings, first proposed by Quispel, Roberts and Thompson, does exist \cite{qrt}. It has the form 
\begin{equation}x_{n+1}={f_1(x_n)-x_{n-1}f_2(x_n)\over f_2(x_n)-x_{n-1}f_3(x_n)}, \label{zdyo}\end{equation}
where the $f_i$ are specific quartic polynomials. The QRT mapping possesses an invariant, which can be expressed in the form of the Euler-Chasles correspondence \cite{veselov}
\begin{equation}\alpha x_n^2x_{n-1}^2+\beta x_nx_{n-1}(x_n+x_{n-1})+\gamma( x_n^2+x_{n-1}^2)+\epsilon  x_nx_{n-1}+\zeta( x_n+x_{n-1})+\mu=0,\label{ztri}\end{equation}
and can be solved in terms of elliptic functions. By the same argument as for the McMillan mapping, one can therefore conclude that all mappings of the QRT family possess the singularity confinement property (unless the invariant \eqref{ztri} is a rational curve, in which case the mapping is linearisable and, as we shall see, the situation becomes more complicated).

Singularity confinement was first discovered, not on mappings, but while studying a fully discrete version of the KdV equation \cite{sincon}, the lattice KdV equation \cite{hirota}. In this chapter however, we choose to focus on the case of mappings and we shall not delve into the properties of integrable lattice equations, which are an even more complicated topic. Suffice it to say that it is the observation that all discrete systems integrable by inverse scattering techniques we studied, in fact possess the singularity confinement property, which led us to propose the latter as a discrete integrability criterion. We shall come back  to this point, but let us first discuss the various types of singularities which may appear in second order rational mappings. In the following we shall always work over  the compactified complex plane, i.e. $x_n\in\mathbb{P}^1$, for all $n$.

Let us perform the singularity analysis (for $a\in\mathbb{C}, a\neq 0$) of the QRT mapping
\begin{equation}x_{n+1}x_{n-1}=a\left(1-{1\over x_n}\right).\label{zpen}\end{equation}
One should first, of course, detect the singularities of the mapping. Remember that a singularity occurs at step $n$ whenever the value of $x_{n+1}$ is independent of $x_{n-1}$ (chosen generically). For \eqref{zpen} this can clearly only happen when $x_n=1$ (yielding $x_{n+1}=0$ for generic $x_{n-1}$) or when $x_n=0$ (yielding $x_{n+1}=\infty$).
Iterating \eqref{zpen} for a generic value $x_0=f$ and for $x_1=1+\epsilon$, and taking the limit $\epsilon\to0$ of all ensuing iterates, it is straightforward to check that one obtains the successive values $0, \infty,\infty, 0$ and $1$, and that the indeterminacy $(1-1)/0$ that arises at $x_7$ is lifted, yielding $x_7=f$, after which the iteration proceeds normally. Hence we have recovered the information on the initial condition $x_0$ that was lost in the singularity at $x_2$ and we conclude that the mapping possesses a {\em confined} singularity pattern $\{1,0,\infty,\infty,0,1\}$. 
However, the second singularity for mapping \eqref{zpen}, resulting from $x_n=0$, is of a very different type. Starting from $x_0=f$ (generic) and $x_1=\epsilon$, we obtain (at the limit $\epsilon\to0$) the sequence of values $f,0,\infty,\infty,0,1/f,\infty,af,0,\infty,\infty,0,1/(af),\infty,a^2f,0,\cdots$, from which it becomes clear that the pattern $\{0,\infty,\infty,0, f', \infty,  f''\}$ of length seven keeps repeating indefinitely. Moreover, it is easy to verify that if we iterate the mapping backwards, we again find the same succession of values, which means that the basic pattern keeps repeating for all $n$. We call such a singularity pattern {\em cyclic}. As will be briefly explained in section \ref{sec-abracadabra}, cyclic patterns are perfectly compatible with the integrable character of a given mapping such as \eqref{zpen}.

As a second example, let us consider the nonintegrable mapping
\begin{equation}x_{n+1}x_{n-1}=1-{1\over x_n^2},\label{zhex}\end{equation}
which has singularities at $x_n=\pm 1$ and $x_n=0$. The first two singularities both behave in exactly the same way: they yield an {\em unconfined}  singularity pattern  $\{\pm1,0,\infty^2,\infty,0^2,\infty^3,\infty^2,0^3,\infty^4,\cdots\}$ which continues indefinitely. (The meaning of the exponents of $\infty$ and 0 is the following: had we introduced a small  quantity $\epsilon$ by assuming that $x_n=\pm1+\epsilon$, we would have found that $x_{n+2}$ is of order $1/\epsilon^2$, $x_{n+4}$ of order $\epsilon^2$ and so on). Note that iterating \eqref{zhex} backwards from $x_n=\pm 1$ and some generic $x_{n-1}=f$, we do not encounter any singularities. This is what sets apart the above unconfined patterns from the singularity pattern obtained for the singularity at $x_n=0$. Iterating \eqref{zhex} forwards from $x_{n-1}=f$ (generic) and $x_n=0$, we obtain a sequence of values similar to that for the unconfined patterns: $\infty^2,\infty,0^2,\infty^3,\infty^2,\cdots$. However, iterating backwards from these initial conditions we find the sequence $\infty, 1/f, 0, \infty^2, \infty, 0^2, \infty^3, \infty^2,\cdots$, showing that in this case the inverse mapping also leads to an unconfined singularity. We thus find what we call an {\em anticonfined} singularity pattern
\begin{equation}
\{\cdots, \infty^4, 0^3, \infty^2, \infty^3, 0^2, \infty, \infty^2, 0 , \frac{1}{f}, \infty, f, 0, \infty^2,\infty,0^2,\infty^3,\infty^2,0^3,\infty^4, \infty^3,\cdots\}
\end{equation}
in which singularities extend indefinitely, both ways, from a finite set of regular values. 

For non-linearisable mappings, the appearance of an unconfined pattern indicates its nonintegrability. Anticonfined patterns on the other hand come in different varieties (as will become clear at the end of this section), some compatible with integrability, some with linearisability and others indicating nonintegrability.

\paragraph{Deautonomising integrable mappings}
The main application of the singularity confinement criterion has been the so-called {\em deautonomisation} \cite{capel} procedure. This procedure consists in deriving integrable, non-autonomous, extensions of integrable autonomous mappings by assuming that the parameters of the latter are functions of the independent variable, the precise form of which is obtained by applying the confinement criterion. We can illustrate this in the case of the mapping (\ref{zpen}), which we rewrite as
\begin{equation}x_{n+1}x_{n-1}=a_n\left(1-{1\over x_n}\right).\label{zhep}\end{equation}
We require that the confined singularity pattern be the same as in the autonomous case and obtain for the function $a(n)$ the constraint
\begin{equation}a_{n+5}a_{n+2}=a_{n+4}a_{n+3},\label{zoct}\end{equation}
the solution of which is $\log a_n=\alpha n+\beta+\gamma(-1)^n$. This non-autonomous form of (\ref{zpen}) was first derived in \cite{papy} where we showed that it is a $q$-discrete form of the Painlev\'e II equation. The cyclic singularity pattern for (\ref{zena}) carries over to the non-autonomous case as well. Starting from the same initial conditions $x_0=f$ and $x_1=0$ we find, using (\ref{zoct}), the succession of values $f,0,\infty,\infty,0,1/f,\infty, a_6 f, 0, \infty, \infty, 0, 1/(a_6 f),\cdots$, i.e. the pattern $\{0,\infty,\infty,0, f', \infty, f''\}$ still repeats indefinitely. (The fact that the cyclic pattern remains cyclic after deautonomisation is not a general feature: in many cases a cyclic pattern becomes a genuinely confined one when deautonomised, see e.g. \cite{cascade}).

The deautonomisation procedure has been instrumental in deriving the non-autonomous forms of most discrete Painlev\'e equations known to date. In fact, it is by this very method that the first $q$-discrete Painlev\'e equations were obtained. Moreover, the structure of the singularity patterns can provide an indication as to where to look for more integrable systems. When studying equations associated to the affine Weyl group E$_8^{(1)}$ (in the Sakai classification \cite{sakai}) we observed that the two previously known equations, obtained by two of the present authors in collaboration with Ohta \cite{ohta}, had singularity patterns of length 7 and 3, and 5 and 5 respectively. Thus we surmised that mappings with singularity patterns of lengths 8 and 2, and 6 and 4 should also exist. This turned out to be indeed the case, allowing us to complement the list of the E$_8^{(1)}$-related discrete Painlev\'e equations \cite{eight}. Also, when deautonomising QRT mappings, the cyclic patterns of the original autonomous mapping yield important information on the geometric structure of the discrete Painlev\'e equations one obtains (as explained in \cite{cascade}). The algebro-geometric underpinnings of the deautonomisation of QRT mappings have been developed in \cite{antonetco}.

\paragraph{Singularities and degree growth}
A most interesting aspect of the singularity structure of rational mappings is that it is intimately related to the degree growth of their iterates. Quoting Veselov, we remind here that ``integrability has an essential correlation with the weak growth of certain characteristics''. The {\em dynamical degree} of a rational mapping is a measure of this growth. It is obtained from the degrees $d_n$ of the iterates of some initial condition and is defined as $\lambda=\lim_{n\to\infty} d_n^{{1/n}}$. Note that $\lambda\geq1$, and that integrable mappings have a dynamical degree equal to 1, while a dynamical degree greater than 1 indicates nonintegrability. (One often encounters an alternative measure of growth, dubbed algebraic entropy \cite{bellon}: it is simply the logarithm of the dynamical degree). In order to illustrate how the singularity structure is linked to the degree growth we consider the McMillan mapping (\ref{zena}). We take initial conditions $x_0=r$ and $x_1=p/q$, and after iterating we find
\begin{gather}
x_2=\frac{r (p^2-q^2) + 2 \mu p q}{q^2-p^2},\qquad x_3= \frac{P_5}{q\, P_2^+ P_2^-}\,,\qquad x_4=\frac{(q^2-p^2)^2\, P_8}{(q^2-p^2)^2\, P_4^+ P_4^-}\,,\label{nonsimp}
\end{gather}
where $P_n, P_n^\pm$ are irreducible polynomials in $p,q,r$ of degree $n$ in $p,q$. We remark that if $q=\pm p$, we have $P_2^\pm\big|_{q^2=p^2}= \pm 2 \mu p^2,\, P_5\big|_{q^2=p^2}=4 \mu^2 p^5$ (and $P_4^\pm\big|_{q^2=p^2}= \pm 4 \mu^2 p^3 q,\, P_8\big|_{q^2=p^2}= 16 \mu^4 p^8 r$), from which we precisely obtain the confined singularity patterns for \eqref{zena}:  $\{\pm1,\infty,\mp1\}$. 
Note that we have written these iterates without enforcing any simplifications, and that they therefore show that the very first cancellation happens in the iterate $x_4$, which is exactly where the indeterminacy occurs that allows the singularities to confine. Clearly, in $x_4$, the degree will drop by four because of the cancellation of $(q^2-p^2)^2$. 
Iterating further we find
\begin{equation}x_5=\frac{(q^2-p^2)^4\,\left(P_2^+ P_2^-\right)^2\, P_{13}}{q (q^2-p^2)^4\, \left(P_2^+ P_2^-\right)^2\, P_6^+ P_6^-},\end{equation}
in which we now have an impressive cancellation (of a polynomial of degree 16).
These massive simplifications, which start appearing at the confinement step and which become increasingly important as one iterates further, have as a result that the degree growth of this mapping is polynomial, rather than exponential (and the mapping thus has a dynamical degree equal to 1). Indeed, computing the degree in $p,q$ of the successive iterates we find the sequence of values, 
$0,1,2,5,8,13,18,25,32,41,50, 61,72,85,98,113,128\cdots$. The degree growth can in fact be rigourously shown to be quadratic: $d_n= \left(n^2+\psi_2(n) \right)/2$, where $\psi_2(n)$ is a periodic function, with period two, defined by $\psi_2(0)=0$ and $\psi_2(1)=1$ (cf. formula \eqref{hd2} in section \ref{sec-rod}).

\paragraph{The question of late confinement}

As explained above, when deautonomising an integrable mapping using singularity confinement, standard practice is to take the patterns obtained for the original autonomous mapping and require that the singularity be confined at the same step as in the autonomous case. Consider for instance the mapping 
\begin{equation}x_{n+1}+x_{n-1}={a_n\over x_n}+{1\over x_n^2}.\label{dhep}\end{equation}
In the autonomous case its (confined) singularity pattern is $\{0,\infty^2,0\}$, which means that if we start from $x_n=f$ (generic) and $x_{n+1}=0$, we recover the information on $f$ at  $x_{n+4}$ (in fact,  $x_{n+4}\equiv f$). Repeating the same procedure when $a$ is a function of $n$ we find that for $x_{n+4}$ to depend on $f$, the function $a_n$ must obey the constraint
\begin{equation}a_{n+3}-2a_{n+2}+a_{n+1}=0,\label{doct}\end{equation}
the solution of which is $a_n=\alpha n+\beta$. This is the well-known deautonomisation of (\ref{dhep}) leading to a discrete form of the Painlev\'e I equation. However, one could also ask what happens if one does not impose the constraint (\ref{doct}). In that case the singularity does not confine at the fourth step but continues along the pattern $\{0,\infty^2,0,\infty, 0,\infty^2,0,\infty,0,\infty^2,0,\cdots\}$. Examining these iterates carefully, it turns out that it is possible for the singularity to be confined at the eighth iterate, provided one imposes the constraint
\begin{equation}a_{n+7}-2a_{n+6}+a_{n+5}-a_{n+4}+a_{n+3}-2a_{n+2}+a_{n+1}=0.\label{denn}\end{equation}
Looking for the solutions of (\ref{denn}) in the form $a_n=\lambda^n$ we obtain the characteristic equation $(\lambda^2-\lambda+1)(\lambda^4-\lambda^3-\lambda^2-\lambda+1)=0$ which leads to solutions for $a_n$ that contain no secular terms. Similar results are obtained if one postpones the confinement even further. Note that, as a rule, such {\em late} confinements always lead to nonintegrable systems \cite{hvdpi,mase1}. 

The case of (\ref{dhep}) is not exceptional. All (at least to the authors' knowledge) integrable mappings, when deautonomised, offer the possibility for late confinement. Thus the rule of thumb when deautonomising with singularity confinement should be to confine at the first occasion, lest a late confinement induce nonintegrability. (A word of caution is necessary here. It may happen in some exceptional cases that a possibility for confinement appears before that which is considered as the timely one. It turns out that in all such cases, such an {\em early} confinement leads to a trivial mapping, usually a periodic one).

\paragraph{Nonintegrable mappings with confined singularities}

The above example shows that at least for late confining mappings, the singularity confinement criterion is not sufficient to ensure integrability, but one could wonder if this a problem that is particular to non-autonomous systems. It turns out that this is not the case. The best known example of a confining but nonintegrable, autonomous mapping is the Hietarinta-Viallet (H-V) mapping \cite{hv}
\begin{equation}x_{n+1}+x_{n-1}=x_n+{1\over x_n^2},\label{ddek}\end{equation}
the confined singularity pattern of which is $\{0,\infty^2,\infty^2,0\}$. The H-V mapping has been extensively studied and its nonintegrability has been rigorously established \cite{takenawaHV}, e.g.: its dynamical degree is $(3+\sqrt 5)/2$. The mapping proposed by Hone \cite{honeviallet}, who attributes it to Viallet, 
\begin{equation}x_{n+1}x_{n-1}=x_n+{1\over x_n},\label{vena}\end{equation}
is another example. Its confined singularity patterns are $\{\pm i,0,\infty,\infty^2,\infty,0,\mp i\}$ and its dynamical degree is greater than 1: its exact value, $(1+\sqrt{17})/4+\sqrt{(1+\sqrt{17})/8}\approx2.0810$, was obtained in \cite{viallethone}. A third example, proposed by Mimura and collaborators \cite{mimura}, is the mapping 
\begin{equation}x_{n+1}x_{n-1}={x_n^4-1\over x_n^4+1},\label{vdyo}\end{equation}
the singularity patterns of which are, $\{\pm1,0\mp1\}$, $\{\pm i,0\mp i\}$, $\{\pm r,0\mp r\}$ and $\{\pm ir,0\mp ir\}$, where $r$ is the square root of $i$, and which has dynamical degree $2+\sqrt 3$. 

It should be clear that infinitely many such examples exist. Moreover, they are not limited to second-order mappings: higher order ones with similar properties do exist and the same is true for lattice equations \cite{kankilattice,latticefulldeauto}. 

As we saw in the preceding paragraphs the confinement of a singularity is intimately linked to cancellations in the iterates of the mapping, which lower its degree growth. For integrable mappings this results in polynomial growth, but while for nonintegrable mappings with confined singularities such cancellations still occur, these turn out to be insufficient and the growth of the mapping remains exponential.

\paragraph{Linearisable mappings and confinement}

Though it might not be a sufficient integrability criterion, the fact that all discrete systems integrable by inverse scattering techniques that we have studied do possess confined singularities, is a strong indication for the necessary character of the confinement property. However, if one relaxes the notion of `integrability' so as to include systems that can be integrated by different means, this conclusion has to be adjusted. For example, it is a well-known  fact that continuous systems that can be considered to be integrable because they are {\em linearisable}, do not necessarily possess the Painlev\'e property \cite{tremblay}. The same turns out to be true for discrete systems. 

As has been thoroughly documented, linearisable second-order mappings belong to one of three classes. The first is that of {\em projective} mappings \cite{karra}. These have the form 
\begin{equation}x_{n+1}x_nx_{n-1}+ax_nx_{n-1}+bx_{n-1}+c=0,\label{vtes}\end{equation}
and can be linearised through a Cole-Hopf transformation $x_n=w_{n+1}/w_n$ to the linear equation $w_{n+2}+a_nw_{n+1}+b_nw_n+c_nw_{n-1}=0$. Projective mappings do have the confinement property, the confined pattern being simply $\{0,\infty\}$.

The second class of linearisable mappings is that of so-called {\em Gambier} mappings \cite{gambier} which consist of two coupled homographic mappings in cascade. One example is $x_{n+1}=y_n+c/x_n$, $\,y_n=a+1/y_{n-1}$ which can be rewritten as
\begin{equation}x_{n+1}=a+{c\over x_n}+{x_{n-1}\over x_nx_{n-1}-c}.\label{vpen}\end{equation}
Equation (\ref{vpen}) has one confined singularity, with pattern $\{0,\infty\}$, but $x_n=\infty$ gives rise to an unconfined one: iterating from this point onwards the memory of $x_{n-1}$ is irretrievably lost. Most Gambier mappings possess unconfined singularities. However, in some cases, when the mapping is sufficiently rich, it is possible to turn these unconfined singularities into confined ones by imposing appropriate constraints on the parameters in the mapping. 

Another interesting Gambier mapping is 
\begin{equation}x_{n+1}=a+x_n-{1\over x_n}+{1\over x_{n-1}},\label{vhex}\end{equation}
obtained by the composition of $x_{n+1}=y_n-1/x_n$ and $y_n=a+y_{n-1}$. This mapping has the confined singularity pattern $\{0,\infty\}$, but also an anticonfined one: $\cdots, 0,0,0,x,\infty, \infty, \infty,\cdots$ (remember that in an anticonfined pattern, singularities extend both ways from a finite set, here a singleton, of regular values). 

The third type of linearisable mapping is known under the moniker of {\em third kind}. These were first discovered in \cite{third} where we have given a general framework for their linearisation. An example of such  a mapping is
\begin{equation}{2z\over x_n+x_{n+1}}+{2z\over x_n+x_{n-1}}=1+{2z\over x_n}.\label{vhep}\end{equation}
It has the confined singularity pattern $\{0,0\}$, but also the anticonfined pattern
\begin{equation}\{\cdots,\infty^2,\infty,\infty,x,x',x'',\infty, \infty,\infty^2, \infty^2,\infty^3,\infty^3,\cdots\}.\label{vhepp}\end{equation}
Notice the linearly growing exponents in the singularities in \eqref{vhepp}, which are the signature of third-kind mappings. The question of whether there exist third-kind autonomous mappings with unconfined singularities remains open.
Deautonomising (\ref{vhep}), based on the pattern $\{0,0\}$, we obtain
\begin{equation}{z_n+z_{n+1}\over x_n+x_{n+1}}+{z_n+z_{n-1}\over x_n+x_{n-1}}=1+{z_{n+1}+z_{n-1}\over x_n},\label{voct}\end{equation}
where $z_n$ is a free function of the independent variable. The linearisation of (\ref{voct}) was presented in \cite{linear}. 
Note that although the deautonomisation preserves the confined pattern, the anticonfined one turns into the unconfined pattern $\{\infty,\infty,\infty,\dots\}$, in which all infinities have exponent 1.

\section{The full-deautonomisation approach}\label{sec-fulldeauto}
All this leads to the question: why is deautonomisation, based on singularity confinement, so successful ? If we leave aside the special case of linearisable mappings the answer is rather obvious: when deautonomising one starts from an autonomous integrable mapping, i.e. a system where the degree growth is slow, associated to given confined and cyclic singularity patterns. As the deautonomisation process preserves the singularity patterns, with the only change that some cyclic patterns may become confined, the simplifications in the iterates that occur in the autonomous case carry over to the non-autonomous one, thus  guaranteeing slow growth and the integrability of the deautonomised system.
However, the deautonomisation approach is even more successful than it would appear at first sight. In fact, when studying the phenomenon of late confinement, we made the remarkable discovery that the confinement conditions on the parameters in the deautonomised mapping, such as \eqref{doct} or \eqref{denn}, actually contain information on the value of the dynamical degree for the resulting non-autonomous mapping.  For example, the largest root of the characteristic equation for \eqref{doct} is clearly 1, the value of the dynamical degree for the associated Painlev\'e I equation, but more strikingly: the largest root for the characteristic equation for \eqref{denn} is approximately 1.72208, which fits exactly with the dynamical degree for the corresponding nonintegrable mapping. The deeper algebro-geometric reasons for this remarkable phenomenon will be briefly touched upon in section \ref{sec-abracadabra}. Here, we shall first explain how, based on these findings, one can extract the value of the dynamical degree for a given mapping, from a `sufficiently general' deautonomisation of that mapping. As will become clear in the following, this approach, dubbed {\em full-deautonomisation} \cite{redemption}, in fact redeems singularity confinement as an integrability criterion.

\paragraph{Deautonomising nonintegrable mappings}
Let us start from mapping (\ref{vena}) which has the confined singularity patterns $\{\pm i,0,\infty,\infty^2,\infty,0,\mp i\}$, and let us deautonomise it as
\begin{equation}x_{n+1}x_{n-1}=x_n-{q_n\over x_n}.\label{tena}\end{equation}
When we require the non-autonomous mapping to possess the same singularity patterns as the autonomous one, we obtain the constraint
\begin{equation}q_{n+7}q_{n+1}=q_{n+6}^2q_{n+2}^2, \label{tdyo}\end{equation}
the largest root of the characteristic equation of which, $(\lambda^2-\lambda+1)(\lambda^4-\lambda^3-2\lambda^2-\lambda+1)=0$, takes precisely the value of the dynamical degree $(1+\sqrt{17})/4+\sqrt{(1+\sqrt{17})/8}$ given in the previous section. 

The mapping (\ref{vdyo}), can be deautonomiseed as
\begin{equation}x_{n+1}x_{n-1}={x_n^4-q_n^4\over x_n^4+1},\label{ttri}\end{equation}
in which form it has the same singularity patterns as in the autonomous case if 
\begin{equation}q_{n+1}q_{n-1}=q_n^4.\label{ttes}\end{equation}
The characteristic equation for this constraint,  $\lambda^2-4\lambda+1=0$, has $2+\sqrt 3$ as its largest solution, which is precisely the value of the dynamical degree for the mapping. 

\paragraph{Full deautonomisation}
Simple deautonomisations such as those for the two previous mappings however do not, in general, suffice to obtain the exact value of the dynamical degree  and often a more sophisticated approach is required. The H-V mapping \eqref{ddek} is a case in point. Let us first try the simple deautonomisation,
\begin{equation}x_{n+1}+x_{n-1}=x_n+{q_n\over x_n^2},\label{tpen}\end{equation}
requiring that the singularity pattern remain the same as for the autonomous case: $\{0,\infty^2,\infty^2,0\}$. Unfortunately, the resulting confinement constraint is just $q_{n+3}=q_n$ and $q_n$ therefore exhibits no growth. However, \eqref{tpen} is not the only possible deautonomisation of the H-V mapping. In particular, one can add a term inversely proportional to $x_n$ and still conserve the singularity pattern of the autonomous mapping:
\begin{equation}x_{n+1}+x_{n-1}=x_n+{f_n\over x_n}+{1\over x_n^2}.\label{thex}\end{equation}
The confinement constraint is now 
\begin{equation}f_{n+3}-2f_{n+2}-2f_{n+1}+f_n=0,\label{thep}\end{equation}
with characteristic equation $(\lambda+1)(\lambda^2-3\lambda+1)=0$, the largest root of which  is precisely the dynamical degree of the H-V mapping: $(3+\sqrt5)/2$. 

We can now define what we mean by `full-deautonomisation'. Ordinarily, when deautonomising a mapping we just assume that its coefficients depend on the independent variable and we fix this dependence by requiring that the singularity patterns of the non-autonomous mapping be the same as that of the autonomous mapping. Full-deautonomisation carries this one step further. Namely, we extend the mapping by adding terms which, though initially absent, do not, when present, modify the singularity patterns and we deautonomise these terms as well.

Let us clarify this approach on a second example,
\begin{equation}x_{n+1}+x_{n-1}={1\over x_n^4},\label{toct}\end{equation}
the confined singularity pattern of which is $\{0,\infty^4,0\}$. Considering all possible terms that leave this singularity pattern unchanged, we find that the extension which leads to an interesting result is, again, through adding a term inversely proportional to $x_n$:
\begin{equation}x_{n+1}+x_{n-1}={a_n\over x_n}+{b_n\over x_n^4}.\label{tenn}\end{equation}
The confinement constraints in this case are $b_{n+1}=b_{n-1}$ (which shows that a simple deautonomisation of (\ref{toct}) does not lead to a helpful result) and
\begin{equation}a_{n+1}-4a_n+a_{n-1}=0.\label{tdek}\end{equation}
This last constraint gives rise to the characteristic equation $\lambda^2-4\lambda+1=0$, the largest root of which is $2+\sqrt3$, coinciding with the dynamical degree of (\ref{toct}).

Lest one get the impression that the full-deautonomisation approach is only applicable to nonintegrable mappings, let us give an example of its use in the case of an integrable system. Let us consider the integrable mapping 
\begin{equation}x_{n+1}+x_{n-1}={1\over x_n^2},\label{qena}\end{equation}
with confined singularity pattern $\{0,\infty^2,0\}$, which we can deautonomise simply by replacing the numerator of the right-hand side by a function $b_n$. Imposing the same singularity pattern we find the constraint $b_{n+1}=b_{n-1}$, which is a trivial non-autonomous extension of (\ref{qena}) since the freedom introduced by the period-2 coefficient $b_n$ can be absorbed in a proper gauge of $x_n$. In the spirit of full-deautonomisation however, it is possible to extend (\ref{qena}) by adding a term proportional to $1/x$ on the right-hand side, while still preserving the singularity pattern. We then obtain the mapping
\begin{equation}x_{n+1}+x_{n-1}={a_n\over x_n}+{1\over x_n^2},\label{qdyo}\end{equation}
which is of course mapping (\ref{dhep}) of section \ref{sec-singconf}, where it was shown that $a_n$ must satisfy the constraint $a_{n+1}-2a_n+a_{n-1}=0$. The corresponding characteristic equation has a double root at $\lambda=1$, which will be seen in section \ref{sec-abracadabra} to be the hallmark of an integrable system.

\paragraph{Late confinement revisited}

In section \ref{sec-singconf} we introduced the notion of late confinement and showed that it leads to nonintegrable systems. In particular, when examining mapping (\ref{dhep}), we found for its first late confinement the constraint (\ref{denn}) with characteristic equation
\begin{equation}(\lambda^2-\lambda+1)(\lambda^4-\lambda^3-\lambda^2-\lambda+1)=0.\label{qtri}\end{equation}
We remarked there that the deautonomisation associated to this late confinement does not give rise to a discrete Painlev\'e equation because the characteristic equation leads to solutions without secular terms. We can now carry this argument one step further since, in the light of the full-deautonomisation approach, we expect the dynamical degree of the late-confined mapping subject to \eqref{denn} to be given, in fact, by the largest root of (\ref{qtri}). The latter turns out to take the value $(1+\sqrt{13}+\sqrt{2}\sqrt{\sqrt{13}-1})/4$, which is approximately 1.7221. In order to verify that this is indeed the value of the dynamical degree of the mapping we can compute its degree growth for a coefficient $a_n$ that satisfies $a_{n+2}=a_{n+1}-a_n$ (which automatically satisfies the constraint (\ref{denn}), as is clear from the factor $(\lambda^2-\lambda+1)$ in \eqref{qtri}). This yields the sequence of degrees $0, 1, 2, 5, 9, 17, 30, 54, 94, 164, 283, 489, 843, 1454,2505,4316$$\,\cdots$, the last entries in which grow as powers of 1.723, in good agreement with the expected value for the dynamical degree. 

Another interesting question we can now address is what happens when one postpones the confinement indefinitely. Clearly, such a situation corresponds to an unconfined singularity. Let us start by considering a confinement which is delayed $k$ times for mapping (\ref{dhep}). The singularity pattern in that case consists of $k$ blocs $\{0,\infty^2,0,\infty\}$ terminated by a $\{0,\infty^2,0\}$ bloc, and the characteristic polynomial for the ensuing confinement constraint is
 \begin{equation}P_k(\lambda)=\left({\lambda^{5k}-1\over\lambda-1}\right)\lambda^3(\lambda P_0(\lambda)-1)+P_0(\lambda),\label{qtes}\end{equation}
where $P_0(\lambda)=\lambda^2-2\lambda+1$ (the characteristic polynomial of the integrable case). Since for $k\geq1$ \eqref{qtes} always has a root greater than 1, we can easily take the limit $k\to\infty$ and we find that the nonconfining version of (\ref{dhep}) (in which $a_n$ does not satisfy condition (\ref{doct}) or any of the late ones), has a dynamical degree given by the largest root of the equation $\lambda P_0(\lambda)-1\equiv\lambda^3-2\lambda^2+\lambda-1=0$, which is approximately $\lambda=1.7549$. Iterating mapping (\ref{dhep}) for arbitrary $a_n$, we obtain the sequence of degrees $0,1,2,5,9,17,30,54,95,168,295,519,911,1600,2808,4929$$\,\cdots$, from which we find that the degree grows approximately as a power of 1.755, which fits quite well with the value we obtained above.
Note that the above degree sequences start to differ at $x_{n+8}$, from which point onwards extra cancellations appear for the late-confined mapping.

The property of late confinement is of course not limited to integrable mappings and one can apply the above approach to confining nonintegrable mappings, using the full-deautonomisation procedure, to obtain their dynamical degree. Several such examples can be found in \cite{redeeming}.

\section{Halburd's exact calculation of the degree growth }\label{sec-rod}

In the previous section we showed that in the full-deautonomisation approach, a simple singularity analysis allows us to decide whether a given mapping is integrable or not and, in the latter case, that it even yields an exact value for the dynamical degree of the mapping. However, applying the full deautonomisation method is not always easy. Although the process of finding and studying all possible extensions of a mapping with confined singularities that do not modify the singularity patterns can be simplified through experience and intuition, in some cases (in particular for higher-order systems) there are substantial difficulties. Thankfully, a simpler method appeared in the ingenious work of Halburd \cite{rodzero}.

Halburd's method starts from the basic fact that the degree of a rational function, $f_n(z)$ say, is equal to the number of preimages of some arbitrary value $w$   for that function, i.e.: the number of solutions  in $z$, counted with the appropriate multiplicity, of $f_n(z)=w$ (in $\mathbb{P}^1$). This is, in fact, the same notion as Arnold's complexity \cite{arnold}. The innovative  feature in Halburd's approach is that this computation of the degree is performed on the $n$-th iterate of a rational mapping, not just for any arbitrary value of $w$, but for values that appear in the singularity patterns of the mapping. 
Starting from Halburd's approach, we presented in \cite{rodone} a simpler method -- dubbed `express' -- which yields the exact value of the dynamical degree of a rational mapping based on its singularity patterns, though not the degree. This information however suffices to decide on the integrability of the mapping, and this in a very efficient way. 

\paragraph{Halburd's method in a nutshell}
In order to calculate the degree of a given mapping we shall view the iterates of the mapping, starting from initial conditions $x_0, x_1\in\mathbb{P}^1$, as rational functions $f_n(z)$ in $x_1=z$ (in which $x_0$ simply appears as a generic constant). Furthermore, as mentioned above, the degree $d_n$ of each iterate will be calculated as the number of solutions, $d_n(w)$, of the equation $f_n(z)=w$, counted with multiplicities, for certain values of $w\in\mathbb{P}^1$. Needless to say that $d_0=0, d_1=1$ and that $d_n=d_n(w)$ for any choice of $w$. Let us show how this calculation works for the McMillan mapping \eqref{zena},
\begin{equation}x_{n+1}+x_{n-1}={2\mu x_n\over 1-x_n^2},\label{hd1}\end{equation}
which has exactly two singularity patterns, both confined: $\{1,\infty,-1\}$ and $\{-1,\infty,1\}$.

We are interested in knowing how many times, at iteration step $n>1$, a particular value of $w$, such as $1$ or $-1$, can appear as the value of $f_n(z)$ for a special choice of $z$. Let us denote the number of appearances of $1$ at step $n$ by $U_n$ and those of $-1$ by $M_n$. Now, a value $1$ can appear either `spontaneously' by an accidental choice of an appropriate $z$, or it arises `automatically' two steps after a value $-1$ appeared for some $z$, as is clear from the singularity pattern $\{-1,\infty,1\}$. The same is true, mutatis mutandis, for the value $-1$. Hence we have that $d_n(1)=U_n+M_{n-2}$ and $d_n(-1)=M_n+U_{n-2}$. What about  the number of possible appearances of $\infty$? Such a value clearly appears automatically one step after a $1$ or $-1$, but can it arise in other circumstances? It is easy to check that the only other possibility for an $\infty$ to appear is two steps after another $\infty$, i.e., as part of a `cyclic pattern' $\{\infty,f\}$. Note that this pattern does not contain any singularities and is therefore, strictly speaking, not a singularity pattern at all. However, it does allows us to conclude that in addition to those generated by values $\pm1$, an additional $\infty$ automatically appears one step out of two. Hence, $d_n(\infty)=U_{n-1}+M_{n-1}+ \psi_2(n)$, where 
\begin{equation}
\psi_2(n) = \frac{1-(-1)^n}{2}.\label{hd2}
\end{equation}
Furthermore, $M_0=U_0=0, M_1=U_1=1$ and, clearly, $M_n=U_n$ for all $n\geq0$. Hence, from $d_n(1)= d_n(-1) = d_n(\infty)$ we find
\begin{equation}
U_n - 2 U_{n-1} + U_{n-2} = \psi_2(n),\label{hd3}
\end{equation}
which has 
\begin{equation}
U_n= \alpha n + \beta + \frac{n^2}{4} - \frac{(-1)^n}{8},\label{hd4}
\end{equation}
as its general solution. The initial conditions for $U_n$ determine the constants, $\alpha=1/2, \beta=1/8$, and the degree $d_n$ of the $n$-th iterate is then obtained, for example, from $d_n=d_n(1)$:
\begin{equation}
d_n=\frac{n^2+\psi_2(n)}{2},\label{degmicmac}
\end{equation}
which fits exactly with the sequence of degrees calculated in section \ref{sec-singconf}, i.e. : $0,1,2,5,8,13,18,25,$ $32,41,50, 61,72,85,98,113,128,\cdots$.

\paragraph{The express method}
The {\em express} method is based on the observation that in order to deduce that the degree of the iterates of a mapping does not grow exponentially, it is in fact not necessary at all to solve recursion relations such as \eqref{hd3} exactly. Since, for the McMillan mapping, it is readily established that the value $\infty$ for an iterate $f_n(z)$ can only occur either because it is induced by a confined singularity pattern, or because it arises in a cyclic pattern, it is clear that the fact that the homogeneous part of equation \eqref{hd3} does not give rise to exponential growth, is sufficient to conclude that the general solution to the full equation cannot exhibit exponential growth. Indeed, the characteristic equation for the homogeneous part of \eqref{hd3}
\begin{equation}
\lambda^2 - 2 \lambda + 1=0,\label{e1}
\end{equation}
lacking any roots greater than 1, and the contribution in the righthand side of \eqref{hd3} due to the occurrences of $\infty$ outside the confined singularity patterns being bounded (in fact, periodic) in $n$, it is clear that the growth of the degree for this mapping can only be polynomial and never exponential. Hence, its dynamical degree is exactly 1 and the mapping is integrable.

This is the crux of the express method: if singularity analysis shows that a given mapping only has confined singularity patterns and cyclic patterns (we will come to the slightly more complicated case of anticonfined patterns in a moment), then it suffices to study the homogeneous parts of the recursion relations that hold for the occurrences of the values that make up the confined singularity patterns. If the resulting characteristic equations do not possess any roots greater than 1, then the mapping at hand is an integrable one. However, if the characteristic equations do possess roots greater than 1, then this is an indication of nonintegrability, as we shall see in a moment. Let us first give two more integrable examples.

In section \ref{sec-singconf} we saw that the mapping (\ref {dhep}) (with $a_n$ satisfying $a_{n+1}-2a_n+a_{n-1}=0$)
\begin{equation}x_{n+1}+x_{n-1}={a_n\over x_n}+{1\over x_n^2},\label{pdyo}\end{equation}
has the (unique) confined singularity pattern $\{0,\infty^2,0\}$. Denoting the number of appearances of the values $0$ and $\infty$ in the iterates of this mapping by $Z_n$ and $I_n$ respectively, it is clear that $d_n(0)=Z_n+ Z_{n-2}$. Moreover, it is easy to establish that in this case as well, an $\infty$ can only appear either due to the confined pattern or in the cyclic pattern $\{\infty,f\}$, the contribution of which we shall neglect and we write $d_n(\infty)\simeq 2 Z_{n-1}$ (where the factor 2 is due to the multiplicty of $\infty$ in the confined pattern and where $\simeq$ signifies equality up a bounded function of $n$). From $d_n(0)=d_n(\infty)$ we then obtain the equation
\begin{equation}
Z_n+Z_{n-2}-2Z_{n-1}\simeq 0,\label{e2}
\end{equation}
the characteristic equation for which, again, has no roots greater than 1, and we conclude  that the mapping is  integrable.

In section \ref{sec-singconf} we also saw that mapping \eqref{zpen} has a confined singularity pattern $\{1,0,\infty,\infty,0,1\}$ and a cyclic pattern $\{0,\infty,\infty,0, f', \infty, f''\}$. Denoting the number of appearances of a value 1 in the iterates of the mapping by $U_n$ and neglecting the cyclic occurrences of the values $0$ and $\infty$, we can therefore write (from $d_n(1)=d_n(0)=d_n(\infty)$) 
\begin{equation}
U_n+U_{n-5} \simeq U_{n-1} + U_{n-4} \simeq U_{n-2} + U_{n-3},\label{e3}
\end{equation}
the characteristic equations
\begin{equation}
(\lambda-1)(\lambda^4-1) = 0,\quad  (\lambda^2-1) (\lambda^3-1) =0,\quad (\lambda-1) (\lambda^2-1) =0,\label{e4}
\end{equation}
for which clearly do not possess any roots that are greater than 1, in accordance with the integrable character of the mapping.

\paragraph{The case of nonintegrable mappings}
Up to now we have only given examples of integrable mappings, but the express method works even better for nonintegrable mappings since it actually yields the exact value of the dynamical degree. A particularly interesting example is that of the Hone-Viallet mapping (\ref{vena}) we already encountered in section 2,
\begin{equation}x_{n+1}x_{n-1}=x_n+{1\over x_n},\label{hena}\end{equation}
which has the confined singularity patterns $\{\pm i, 0,\infty,\infty^2,\infty,0,\mp i\}$. This mapping also has a long cyclic pattern involving the values $0$ and $\infty$ but not $\pm i$: $\{f, 0,\infty,\infty^2,\infty,0,f',\infty,\infty\}$. (Note that the confined singularity patterns were not correctly identified in \cite{honeviallet, redeeming}). 

Denoting by $P_n$ ($M_n$) the number of appearances of $+ i$ ($-i$) at the $n^{\rm th}$ iterate, and neglecting the contributions from the cyclic pattern, we find from the various possible expressions for the degrees ($d_n(i)=d_n(-i)=d_n(0)=d_n(\infty)$) that
\begin{align}P_n + M_{n-6} = M_n + P_{n-6} &\simeq (P_{n-1}+M_{n-1}) + (P_{n-5}+M_{n-5})\nonumber\\
&\simeq (P_{n-2}+M_{n-2}) + 2(P_{n-3}+M_{n-3})+(P_{n-4}+M_{n-4}).\label{hdyo}
\end{align}
Now, since $P_n$ and $M_n$ clearly play the same role, we can take $M_n=P_n$ and we find 
\begin{gather}(\lambda^2-\lambda+1) Q(\lambda) = 0,\qquad (\lambda^2+\lambda+1)Q(\lambda) = 0,\qquad Q(\lambda) =0,
\end{gather}
as characteristic equations for \eqref{hdyo}, where $Q(\lambda) = (\lambda^4-\lambda^3-2\lambda^2-\lambda+1)$. The solution to the general recursion relations that can be established for the degree $d_n$ of the iterates of this mapping, can therefore be described in terms of the roots of the polynomial $Q(\lambda)$, among which $(1+\sqrt{17})/4+\sqrt{(1+\sqrt{17})/8}$ is the largest one. In \cite{viallethone} it has been shown that this largest root of $Q(\lambda)$, in fact, gives the dynamical degree for this mapping.

Another interesting nonintegrable example is that of a late confined one. As such we choose the first late confinement of (\ref{dhep}) corresponding to the pattern $\{0,\infty^2,0,\infty, 0,\infty^2,0\}$.
Denoting by $Z_n$ the number of appearances of the value 0 at some iterate $n$ and neglecting all cyclic patterns that might arise, we find from $d_n(0)=d_n(\infty)$ the expression
\begin{equation}Z_n+Z_{n-2}+Z_{n-4}+Z_{n-6}\simeq 2Z_{n-1}+Z_{n-3}+2Z_{n-5},\label{hhex}\end{equation}
the  characteristic equation for which is:
\begin{equation}(\lambda^2-\lambda+1)(\lambda^4-\lambda^3-\lambda^2-\lambda+1)=0.\label{hhep}\end{equation}
This equation is identical to \eqref{qtri} and its largest root, $(1+\sqrt{13}+\sqrt{2}\sqrt{\sqrt{13}-1})/4$, therefore coincides with the value of the dynamical degree obtained by the full-deautonomisation approach.

Many more interesting examples can be found in \cite{rodone}, where it is also explained that if the confined singularity patterns are too short, a straightforward application of the express method might not lead to useful conclusions, and it is shown how one can deal with such a situation. 

\paragraph{Non confining mappings}
The fact that the express method can be applied to cases with late confinement raises the possibility that the approach might also be applicable to mappings with unconfined singularities. This turns out to be indeed the case if the unconfined singularity patterns, although extending indefinitely, are made up of blocks that keep repeating. 
To illustrate this we consider  the mapping
\begin{equation}x_{n+1}+x_{n-1}=x_n+{1\over x_n^3},\label{hoct}\end{equation}
which has the unconfined singularity $\{0,\infty^3,\infty^3,0,\infty^3,\infty^3,0,\cdots\}$.
Denoting by $Z_n$ the number of appearances of the value 0, we have for the preimages of $0$ and $\infty$,
\begin{equation}d_n(0)=Z_n+Z_{n-3}+Z_{n-6}+\cdots=\sum_{k=0}^{\infty} Z_{n-3k},\quad d_n(\infty)=3\sum_{k=0}^{\infty} (Z_{n-3k-1}+Z_{n-3k-2}), \label{henn}
\end{equation}
where $Z_{m}=0$ whenever $m\leq0$ because of the initial conditions we imposed for the iteration, and all sums in these expressions are therefore finite ones.
As a result we obtain the relation
\begin{equation}\sum_{k=0}^{\infty} Z_{n-3k}-3\sum_{k=0}^{\infty} (Z_{n-3k-1}+Z_{n-3k-2})\simeq0,\label{hdek}\end{equation}
and its corresponding characteristic equation by setting $Z_n=\lambda^n$. Assuming now that this characteristic equation has a root with modulus greater than 1, we can take the limit $n\to\infty$ to obtain 
\begin{equation}{1\over1-{1\over\lambda^3}}=\left({1\over\lambda}+{1\over\lambda^2}\right){3\over1-{1\over\lambda^3}},\label{oena}\end{equation}
which yields, 
\begin{equation}\lambda^2-3\lambda-3=0.\label{odyo}\end{equation}
The largest root of (\ref{odyo}) is $\lambda=(3+\sqrt{21})/2$, approximately equal to 3.791. This value of the dynamical degree is in perfect agreement with the one obtained by the direct calculation of the sequence of degrees: 0, 1, 4, 15, 58, 220, 834, 3163,$\cdots$.

\paragraph{Linearisable mappings}
Linearisable mappings are special in the sense that they are integrable without necessarily having confined singularities. Still, their dynamical degree is always equal to 1 and the degree of their iterates grows linearly with $n$. Applying the express method to linearisable mappings is straightforward. Let us consider the mapping
\begin{equation}{x_{n+1}+x_n\over x_{n-1}+x_n}={1-x_{n}\over 1+x_{n}},\label{otri}\end{equation}
which can be linearized as $y_n=y_{n-1}+1$, $x_{n+1}=(1-x_ny_n)/(y_n-1)$ \cite{nonqrt}. The mapping has two singularity patterns, a confined one, $\{1,-1\}$, and an unconfined one, $\{-1,\infty,\infty,\infty,\dots\}$, and it is easily checked that, starting from $x_0=r, x_1=p/q$, the growth of the degree (in $p,q$) of its iterates is linear: 0, 1, 2, 3, 4, 5, 6, 7, 8, 9, \dots. 

We denote by $U_n$ and $M_n$ the number of appearances of the values $+1$ and $-1$, respectively, at the $n^{\rm th}$ iterate of the mapping. From the preimages of the values $+1, -1$  and $\infty$ we find that 
\begin{equation}U_n= M_n+U_{n-1}\simeq M_{n-1}+M_{n-2}+M_{n-3}+\dots=\sum_{k=1}^{\infty}M_{n-k},\label{otes}\end{equation} 
with $U_m=M_m=0$ if $m\leq 0$, in which we substitute $U_n=U_0\lambda^n$ and $M_n=M_0\lambda^n$ in order to obtain a characteristic equation for which we assume again that it has a root greater than 1 so that we can  resum the geometric series that arises at the limit $n\to\infty$. We obtain finally
\begin{equation}
M_0=U_0\left(1-{1\over\lambda}\right)
\quad{\rm and}\quad
U_0 = M_0 \left(\frac{1}{1-1/\lambda}-1\right),\label{ohep}\end{equation}
and find $\lambda=1$ as the only possible solution. This means that the root $\lambda > 1$ that we assumed cannot exist. However, as by definition the dynamical degree cannot be less than 1, it then follows that the dynamical degree of (\ref{otri}) must be equal to 1, as expected.

Of course, not all linearisable mappings have unconfined singularities. For instance, the mapping (linearised in \cite{mimuralin})
\begin{equation}x_{n+1}x_{n-1}=x_n^2-1,\label{ooct}\end{equation}
has two confined singularity patterns $\{\pm1,0,\mp1\}$ and an anticonfined one
\begin{equation}
\cdots, \infty^4, \infty^3, \infty^2, \infty , f, 0, f', \infty, \infty^2, \infty^3, \infty^4,\cdots\label{QRTanticonf}. 
\end{equation}
The degree of its iterates (for $x_0=r, x_1=p/q$) grows linearly as $0,1,2,4,6,8,10,12\hdots$, and it belongs to the class that we call `linearisable of the third kind'. From these singularity patterns it is immediately clear that, if we denote by $U_n, M_n$ the number of appearances of the values $+1, -1$  in the iteration, we have that $d_n(1) = U_n +M_{n-2}$, $d_n(-1)=M_n+U_{n-2}$ and $d_n(0)=U_{n-1} + M_{n-1} + \delta_{n1}$ (where the Kronecker $\delta$ indicates the single appearance of a 0 in the anticonfining pattern). Taking $M_n=U_n$ (since +1 and -1 obviously play the same role) and neglecting  -- just as we did for the cyclic patterns -- the contribution of the anticonfined pattern, we have from $d_n(1)=d_n(0)$ that
\begin{equation}U_n+U_{n-2}\simeq2 U_{n-1}.\label{oenn}\end{equation}
As the characteristic equation for this relation does not have any roots greater than 1, we find, as expected, that the dynamical degree for this mapping is 1.

\paragraph{The effect of anticonfinement}

The previous example suggests that there might exist cases where one cannot just brush aside the anticonfined singularities. The mapping introduced by Tsuda and collaborators \cite{tsuda} is a case in point:
\begin{equation}x_{n+1}=x_{n-1}\left(x_n-{1\over x_n}\right).\label{ndyo}\end{equation}
This nonintegrable mapping has  two confined singularity patterns, $\{\pm1,0,\infty,\mp1\}$, as well as an anticonfined one
\begin{equation}\{\cdots, 0^8,0^5,0^3,0^2,0,0,f,0,\infty,f', \infty,\infty,\infty^2,\infty^3,\infty^5,\infty^8\cdots\},\end{equation}
in which the exponents form a Fibonacci sequence. Due to this exponential growth of the number of occurrences of the value $\infty$ in the iterates of the mapping, one concludes that its dynamical degree must be at least equal to the golden mean $\varphi=(1+\sqrt 5)/2\approx 1.6180$. Direct calculation of the degree, 0, 1, 2, 4, 8, 14, 24, 40, 66, 108, 176, 286, 464, 752, 1218, 1972, 3192$\dots$, indicates that the growth rate indeed converges to the golden mean.

Denoting by $U_n$ the number of spontaneous occurrences of the value 1 in the iteration (and taking into account that $+1$ and $-1$ play the same role)  we find that $d_n(\pm1)=U_n+U_{n-3}$, $d_n(0)=2U_{n-1}+\delta_{n1}$ where the  $\delta_{n1}$ is due to the single appearance of 0, after a generic value, in the anticonfined pattern. We thus find the equation
\begin{equation}U_n+U_{n-3}=2U_{n-1}+\delta_{n1},\label{npen}\end{equation} 
and the dynamical degree of the mapping, given by the largest root of the characteristic equation for (\ref{npen}),  is precisely the golden mean already obtained above. 

The interesting point here is that, had we tried to compute the degree of the mapping from the number of appearances of the value $\infty$, we would have found $d_n(\infty) = 2U_{n-2}+(f_{n-3}+\delta_{n2})+f_{n}$ (due to the two generic values for $f$ that are followed by an infinite sequence of infinities with Fibonacci exponents in the anticonfined pattern) as an alternative right-hand side for \eqref{npen}. Here, $f_n$ is defined by $f_{n+1}=f_n+f_{n-1} ~\text{for}~ n\ge1$ with $f_1=1$ and $f_n=0$ for $n\le0$, and this type of contribution to the equations cannot simply be discarded in an express-type treatment.
In general, when one has an anticonfined singularity pattern with exponential growth in its exponents, it may happen that this growth coincides with the dynamical degree of the mapping, but there are cases where it only offers a lower bound for the dynamical degree. However, as we have seen above, anticonfined singularities with non-exponential (or even no) growth do exist for some mappings, and the corresponding patterns can be neglected, just as the cyclic ones, when calculating the dynamical degree with the express method. 

\section{Singularities and spaces of initial conditions}\label{sec-abracadabra}

In this section we shall briefly explain the algebro-geometric background of the singularity confinement property. More precisely, we will show on an example that for a second order (bi-) rational mapping that only possesses confined and cyclic singularities, its points of indeterminacy can be resolved by means of a finite number of blow-ups, i.e.: that the mapping is birationally equivalent to a family of isomorphisms between rational surfaces (in the general, nonautonomous, case), obtained by blowing up $\mathbb{P}^1\times\mathbb{P}^1$ a finite number of times. In the process, it will also become clear why the deautonomisation and, especially, the full-deautonomisation procedures work in the way they do.

As an example we choose the nonautonomous mapping \eqref{zhep}, for $a_n \ne 0$, which we rewrite as the birational mapping $\varphi_n \colon \mathbb{P}^1 \times \mathbb{P}^1 \dashrightarrow \mathbb{P}^1 \times \mathbb{P}^1$ :
\begin{equation}
	\left( x_n, y_n \right) \mapsto \left( x_{n+1}, y_{n+1} \right) = \left( y_n, \frac{a_n (y_n - 1)}{x_n y_n} \right).\label{fien}
\end{equation}
As usual, we cover $\mathbb{P}^1 \times \mathbb{P}^1$ with four copies of $\mathbb{C}^2$, as $\mathbb{P}^1 \times \mathbb{P}^1 = (x_n, y_n) \cup (x_n, t_n) \cup (s_n, y_n) \cup (s_n, t_n)$, where $s_n := 1 / x_n$ and $t_n := 1 / y_n$. From the definition \eqref{fien} it is clear that the mapping is indeterminate at the points $(x_n, y_n) = (\infty, 0)$ and $(x_n, y_n) = (0, 1)$ (the so-called points of indeterminacy of $\varphi_n$), and that there exist exactly two curves in $\mathbb{P}^1 \times \mathbb{P}^1$ that contract to a point under its action, the curves $\{ y_n = 1 \}$ and $\{ y_n = 0 \}$:
\begin{equation}
\{ y_n = 1 \} \to (x_{n+1}, y_{n+1}) = (1, 0) ,\qquad
\{ y_n = 0 \} \to (x_{n+1}, y_{n+1}) = (0, \infty).\label{contractedcurves}
\end{equation}
Note that the images of these two curves are nothing but the points of indeterminacy for the inverse mapping $\varphi_n^{-1}$. These are the singularities of the mapping \eqref{fien}.
Moreover, it is easily checked that, under the action of $\varphi_n$, both contracted curves end up in the point of indeterminacy $(\infty,0)$ through the chains
\begin{equation}
\{y_n=1\} \to (1,0) \to (0,\infty) \to (\infty,\infty) \to (\infty,0), \label{confprechain}
\end{equation}
and
\begin{equation}
\{y_n=0\} \to (0,\infty) \to (\infty,\infty) \to (\infty,0),\label{cyclicprechain}
\end{equation}
which clearly correspond to the first parts of the confined singularity pattern $\{1,0,\infty,\infty,0,1\}$ and the cyclic pattern $\{0,\infty,\infty,0,f',\infty,f''\}$ for \eqref{zhep}, respectively. Recall that in the singularity confinement approach the indeterminacy was lifted using a continuity argument in the initial conditions. Here we shall show that the same aim can be achieved by 8 successive blow-ups of $\mathbb{P}^1 \times \mathbb{P}^1$. For this purpose, let us define the points $P_n \colon (x_n, y_n) = (1, 0)$, $Q_n \colon (x_n, y_n) = (0, \infty)$, $R_n \colon (x_n, y_n) = (0, 1)$ and $S_n \colon (x_n, y_n) = (\infty, 0)$, for which it should be remarked that we have $\varphi_{n+1} \circ \varphi_n (Q_n) = S_{n+2}$ for all $n$ and for any choice of $a_n$. Hence, it is clear that we will have achieved confinement, along the lines of the pattern $\{1,0,\infty,\infty,0,1\}$ for mapping \eqref{zhep}, if we can establish that $
\varphi_{n+4}\circ\varphi_{n+3}\circ\varphi_{n+2}\circ\varphi_{n+1} (P_{n+1}) = R_{n+5}.$

In general, we shall say that the singularity for mapping \eqref{fien} that arises at $P_{n+1}$, due to the collapse of the curve $\{y_{n}=1\}$, is confined if there exists some value $k$ for which we have that
\begin{equation}
\varphi_{n+k+1} \circ\,\cdots \circ\,\varphi_{n+2}\circ\, \varphi_{n+1} (P_{n+1}) = R_{n+k+2}.\label{genconf}
\end{equation}

\begin{figure}[t]
\begin{center}
\resizebox{8cm}{!}{\includegraphics{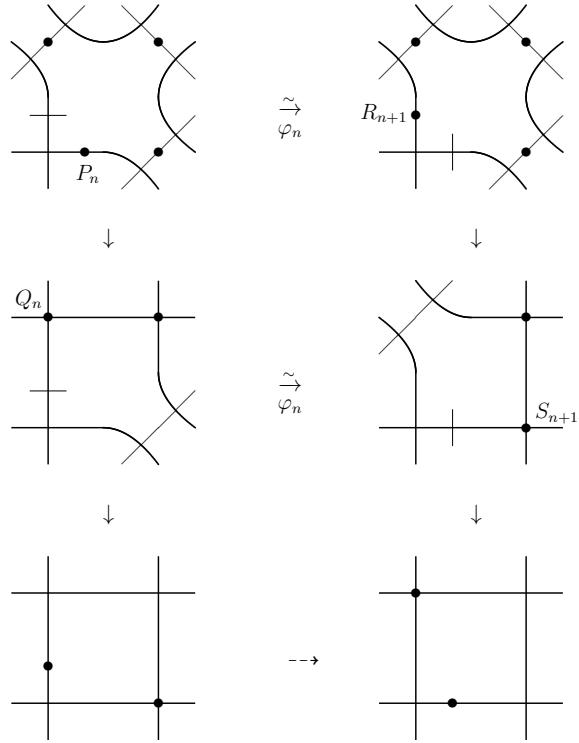}}
\caption{Pictorial representation of the 8 blow-ups of $\mathbb{P}^1 \times \mathbb{P}^1$ that are required to lift the indeterminacy for mapping \eqref{fien} under condition \eqref{timelyconf}.}
\label{fig1}
\end{center}
\end{figure}

\begin{figure}[t]
\begin{center}
\resizebox{6.5cm}{!}{\includegraphics{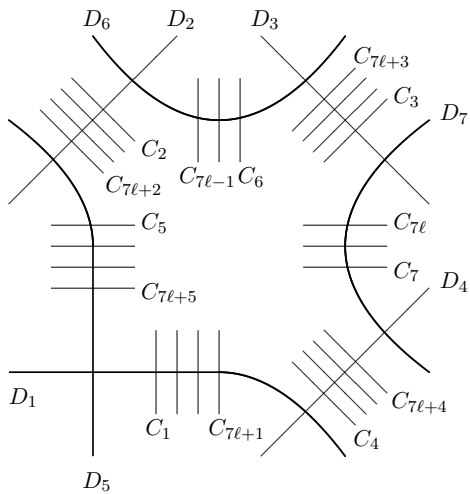}}
\caption{Pictorial representation of the rational surface obtained for mapping \eqref{fien} by blowing up $\mathbb{P}^1 \times \mathbb{P}^1$ at $7\ell+8$ points. The surface for the standard confinement -- which is the same as for the autonomous mapping -- corresponds to $\ell=0$ (in which case curves $C_6$ and $C_7$ are absent).}
\label{fig2}
\end{center}
\end{figure}

\paragraph{Standard confinement} Let us first see  how this can be achieved for the standard confinement (or what we called, in section 2, the `timely' confinement) which arises at $k=3$.
More precisely, we want to show that 
\begin{equation}
\varphi_{n+4}\circ \varphi_{n+3}\circ \varphi_{n+2}\circ \varphi_{n+1} (P_{n+1}) = R_{n+5}~~
	\Leftrightarrow\quad
	\frac{a_{n+1} a_{n+4}}{a_{n+2} a_{n+3}} = 1.\label{timelyconf}
\end{equation}
This condition on the parameters $a_n$, which is automatically satisfied if $a_n$ is independent of $n$ (and the confinement is therefore indeed timely, i.e. the same as in the autonomous case) is of course nothing but the condition \eqref{zoct} we obtained from the deautonomisation approach in section 2.
In order to prove condition \eqref{timelyconf} we have to blow up $\mathbb{P}^1 \times \mathbb{P}^1$, for general $n$, at the 8 points, 
\begin{subequations}
\begin{gather}
\Big( x_n, y_n \Big) = (0, 1) ,\quad (s_n, y_n) = (0, 0) ,\quad  (x_n, t_n) = (0, 0) ,\quad (x_n, y_n) = (1, 0)\label{eightpointsstart}\\ 
\Big( \frac{x_n}{t_n}, t_n \Big) = (-a_{n-1}, 0) ,\quad (s_n, t_n) = (0, 0) ,
\quad \Big( s_n, \frac{t_n}{s_n} \Big) = \Big( 0, -\frac{a_{n-2}}{a_{n-1}} \Big) ,\\
\quad \text{and}\qquad\Big( \frac{s_n}{y_n}, y_n \Big) = \Big( -\frac{a_{n-3}}{a_{n-2} a_{n-1}}, 0 \Big),\label{eightpointsend}
\end{gather}
\end{subequations}
as described in Figure \ref{fig1}. The rational surface $X_n$ (at each value of $n$) obtained from these blow-ups of $\mathbb{P}^1 \times \mathbb{P}^1$ corresponds to the case $\ell=0$ of the surface depicted in Figure \ref{fig2}. 

It can be verified, subject to the condition on the parameters given in \eqref{timelyconf}, that $\varphi_n$ is now defined on the whole of $X_n$ (for every $n$) and that it acts as an isomorphism from $X_n$ to $X_{n+1}$ (and as an automorphism on $X=X_n$ in the autonomous case). The surface $X_n$ is called the space of initial conditions for the mapping, a notion which was first introduced by Okamoto in his study of the differential Painlev\'e equations \cite{okamoto}.

It is easy to check that the curves $D_1, \hdots, D_7$ and $C_1, C_2, C_3$ on $X_n$, as defined in Figure \ref{fig2}, move in the following fashion under the action of the mapping:
\begin{gather}
D_1 \to D_2 \to \cdots \to D_7 \to D_1\quad\text{and}\qquad
\{ y = 1 \} \to C_1 \to C_2 \to \cdots \to C_5 \to \{ x = 1 \}.\label{motion}
\end{gather}
The motion of the $D$ curves clearly corresponds to the cyclic pattern $\{ 0, \infty, \infty, 0, f, \infty, f' \}$, and that of the $C$ curves to the confined pattern $\{ 1, 0, \infty, \infty, 0, 1 \}$ for \eqref{zhep}. As the curves $D_1, \ldots, D_7, C_1, C_2, C_3$ form a basis for the Picard lattice (of rank 10) for $X_n$, \eqref{motion} suffices to define the automorphism $\varphi_{*}$ that is induced by $\varphi_n$ on this Picard lattice, for all $n$. In particular, $\varphi_{*}$ can be represented as the linear map $
\begin{bmatrix}
	U & * \\
	O & \Phi
\end{bmatrix}$,
where $U$ is the permutation matrix of size $7$ that corresponds to the motion of the $D$ curves, and
$\Phi = \begin{bmatrix}
	0 & 0 & -1 \\
	1 & 0 & 1 \\
	0 & 1 & 1
\end{bmatrix}$.
Here $O$ stands for a size $3\times7$ null matrix and we have omitted the entries in the upper right-hand corner of the matrix for $\varphi_*$ as these are irrelevant when it comes to deciding whether the mapping is integrable or not. The reason for this is that, according to \cite{takenawaHV} and \cite{dillerfavre}, the dynamical degree of the mapping actually coincides with the largest eigenvalue for $\varphi_{*}$, which is obviously decided by the submatrices $U$ and $\Phi$.
Since $U$ is a permutation matrix and is thus unitary, its eigenvalues all have modulus $1$.
Moreover, from the form of the submatrix $\Phi$ (which is a companion matrix) it is clear that its eigenvalues are $1, 1$ and $-1$.
Hence, all eigenvalues of the linear action $\varphi_{*}$ have modulus $1$ and the dynamical degree of the mapping, under the constraint
$a_{n} a_{n+3}=a_{n+1} a_{n+2}$,
is necessarily equal to $1$ as well and the mapping is therefore integrable.
Note that this constraint can be written as $\big[A_{n+1}~ A_{n+2} ~A_{n+3}\big]= \big[A_{n}~ A_{n+1} ~A_{n+2}\big]\cdot\Phi$,
where $A_n = \log a_n$. The statement that the submatrix $\Phi$ is free of eigenvalues with modulus greater than 1 is therefore equivalent to saying that the characteristic equation for the confinement constraint on the parameters $a_n$ in the mapping does not have any roots with modulus greater than 1. This is the fundamental observation that underlies the full-deautonomisation approach, as explained in section 2. 

Another important observation is that the minimal polynomial for the submatrix $\Phi$ contains a factor $(\lambda-1)^2$, and that the parameters which satisfy the confinement condition therefore have secular dependence on $n$. This is, in fact, a general feature: the non-permutation part of the linear action $\varphi_*$ obtained by regularising a confining second order mapping by blow-up, can only have a double root $1$ in its minimal polynomial if the mapping is integrable. 

A last remark concerns the type of discrete Painlev\'e equation, in Sakai's classification \cite{sakai}, that we obtain from condition \eqref{timelyconf}.
As is easily established from Figure \ref{fig2}, the intersection diagram of the $D$ curves that make up the cyclic pattern for this mapping (i.e. the diagram obtained by associating a vertex with each curve $D_i$ and an edge between two vertices with each intersection of curves) is nothing but the Dynkin diagram for A$_6^{(1)}$, which is exactly the type of surface in Sakai's clasification that is associated with this particular discrete Painlev\'e equation. Hence, it is clear that the cyclic singularity pattern for the original autonomous mapping \eqref{zpen} in fact determines the surface type of its (standard) deautonomisation (see \cite{cascade} for more details).

\paragraph{Late confinement}
A systematic analysis of condition \eqref{genconf} shows that all late confinements for the mapping \eqref{fien} occur at values $k=7\ell+3$, for some non-negative integer $\ell$. 
To see this, let us introduce a coordinate on each $D_i$ by defining:
\begin{subequations}
\begin{gather}
T^{(1)}_n(\alpha) \colon (s_n, y_n) = (\alpha, 0) ,\quad T^{(2)}_n(\beta) \colon \left( -\frac{x_n}{t_n}, t_n \right) = (\beta, 0) ,\label{sevenpointsstart}\\
T^{(3)}_n(\gamma) \colon \left( s_n, -\frac{t_n}{s_n} \right) = (0, \gamma) ,\quad T^{(4)}_n(\delta) \colon \left( -\frac{s_n}{y_n}, y_n \right) = (\delta, 0) ,\\
T^{(5)}_n(\epsilon) \colon (x_n, y_n) = (0, \epsilon) ,\quad T^{(6)}_n(\zeta) \colon (x_n, t_n) = (\zeta, 0) ,\quad
T^{(7)}_n(\eta) \colon (s_n, t_n) = (\eta, 0).\label{sevenpointsend}
\end{gather}
\end{subequations}
Note that in this notation $T^{(1)}_n(1) = P_n$ and $T^{(5)}_n(1) = R_n$. 
Moreover, we have that
\begin{subequations}
\begin{gather}
\varphi_n(T_n^{(1)}(\alpha)) = T^{(2)}_{n+1}(a_n \alpha) ,\quad\varphi_n(T_n^{(2)}(\beta)) = T^{(3)}_{n+1}(\beta / a_n) ,\label{genconfchainstart}\\
\varphi_n(T_n^{(3)}(\gamma)) = T^{(4)}_{n+1}(\gamma / a_n) ,\quad\varphi_n(T_n^{(4)}(\delta)) = T^{(5)}_{n+1}(a_n \delta) ,\\
\varphi_n(T_n^{(5)}(\epsilon)) = T^{(6)}_{n+1}(\epsilon), \quad\varphi_n(T_n^{(6)}(\zeta)) = T^{(7)}_{n+1}(\zeta / a_n) ,\quad \varphi_n(T_n^{(7)}(\eta)) = T^{(1)}_{n+1}(\eta).\label{genconfchainend}
\end{gather}
\end{subequations}
Now, since
\begin{equation}
	\varphi_{n+7\ell+3}\circ \cdots \circ\varphi_{n+1}\circ \varphi_n (P_n) = T^{(5)}_n \left( \frac{a_{n+7\ell} a_{n+7\ell+3}}{a_{n+7\ell+1} a_{n+7\ell+2}} \prod^{\ell-1}_{m=0} \frac{a_{n+7m} a_{n+7m+3}}{a_{n+7m+1} a_{n+7m+2} a_{n+7m+5}} \right),
\end{equation}
we find that confinement can only happen through condition \eqref{genconf} and this if and only if $k=7\ell+3$, and 
\begin{equation}
	\frac{a_{n+7\ell+1} a_{n+7\ell+4}}{a_{n+7\ell+2} a_{n+7\ell+3}} \prod^{\ell-1}_{m=0} \frac{a_{n+7m+1} a_{n+7m+4}}{a_{n+7m+2} a_{n+7m+3} a_{n+7m+6}} = 1.\label{genlatecond}
\end {equation}
Note that when $\ell\geq1$, this condition is not satisfied for $a_n$ which are independent of $n$. For such late confinements,
blowing up $\mathbb{P}^1\times\mathbb{P}^1$ at the 8 points (\ref{eightpointsstart}--\ref{eightpointsend}) as well as at the $7\ell$ points (\ref{sevenpointsstart}--\ref{sevenpointsend}) that appear in the chain (\ref{genconfchainstart}--\ref{genconfchainend}) before confinement happens, we obtain the rational surface depicted in Figure \ref{fig2} for general $\ell$. On this surface, the $D$ curves still form a cycle of length 7 
\begin{equation}
D_1 \to D_2 \to \cdots \to D_7 \to D_1 ,
\end{equation}
(which proves that in this case the cyclic pattern not only survives the standard deautonomisation but also any late one), but the confined pattern now becomes 
\begin{equation}
\{ y = 1 \} \to C_1 \to C_2 \to \cdots \to C_{7\ell+5} \to \{ x = 1 \},
\end{equation}
or $\{ 1, 0, \infty, \infty, 0, \alpha, \infty, \alpha', \ldots, 0, \infty, \infty, 0, 1  \}$ in the language of mapping \eqref{zhep} (where the entries $\alpha, \alpha'$ etc. take specific values that depend on $a_n$ but not on the initial condition).

The Picard lattice for this surface now has rank $7 \ell + 10$ and we can use the curves $D_1, \ldots, D_7, $ $C_1, \ldots, C_{7\ell+3}$ as a basis. We then find that the linear action $\varphi_*$ induced by the mapping \eqref{fien} on this Picard lattice takes the form
$\varphi_{*} \sim
\begin{bmatrix}
	U & * \\
	O & \Phi
\end{bmatrix}$,
where $U$ is the same as in the standard confinement case and $\Phi$ is again a companion matrix, now of size $7\ell+3$, the last column of which is ${}^t \! \begin{bmatrix}
		-1 & 1 & 1 & -1 & 0 & 1 & 0 & \cdots & -1 & 1 & 1
	\end{bmatrix}
$. The minimal polynomial of $\Phi$ is therefore
\begin{equation}
	\lambda^{7\ell+3} - \lambda^{7\ell+2} - \lambda^{7\ell+1} +  \lambda^{7\ell} - (\lambda^5 - \lambda^3 + \lambda^2 + \lambda - 1) \sum^{\ell-1}_{m=0} \lambda^{7m},
\end{equation}
which always has a real root greater than $1$ unless $\ell = 0$.
Hence, when $\ell\geq1$, the linear action $\varphi_{*}$ has an eigenvalue greater than $1$ and we conclude that the mapping in the late confinement case is always nonintegrable.


\paragraph{Late confinement and full-deautonomisation} Just as for the standard confinement, it is easy to check that the general confinement constraint \eqref{genlatecond} can be 
written as
\begin{equation}
	\begin{bmatrix}
		A_{n+2} & \cdots & A_{n+7\ell+4}
	\end{bmatrix}
	= \begin{bmatrix}
		A_{n+1} & \cdots & A_{n+7\ell+3}
	\end{bmatrix}\cdot
	\Phi,
\end{equation}
where $A_n = \log a_n$.
Now, since the evolution of the coefficient $a_n$ is written in terms of the submatrix $\Phi$, the growth rate of $A_n = \log a_n$ is determined by the largest root of $\varphi_{*}$, which explains why the full-deautonomisation method works in the case of general late confinements as well.
Note that while the full-deautonomisation method has a clear algebro-geometric justification only in the case of second-order mappings, several encouraging results do suggest however that it might be valid for higher order mappings as well \cite{latticefulldeauto}. 

\paragraph{Spaces of initial conditions and degree growth}
The fact that autonomous second order rational mappings that only have confined or cyclic singularities, in fact,  possess a space of initial conditions (in the sense explained above) was first recognized by Takenawa \cite{takenawaHV}. The conditions under which a nonautonomous mapping can be said to enjoy the same property are set out in detail in \cite{mase}.  

In \cite{dillerfavre} it is shown that if an autonomous second order mapping has a space of initial conditions, then its degree growth must be either bounded, quadratic or exponential. Moreover, in the case of exponential growth, the value of the dynamical degree is strongly restricted: it can only be a reciprocal quadratic integer greater than 1 or a Salem number (i.e., a real algebraic integer greater than 1, such that its reciprocal is a conjugate and all (but at least one) of the other conjugates lie on the unit circle). Note that this result implies, in particular, that if a mapping has linear degree growth, it cannot possess a space of initial conditions. A similar result has been shown to hold for nonautonomous mappings as well \cite{mase}. 

If a mapping does not have a space of initial conditions, there is no general theory in the nonautonomous case. In the autonomous case however, it is known that such a mapping can only have linear or exponential degree growth.
Moreover, again in the autonomous case, it is sometimes possible to verify whether or not a mapping has a space of initial conditions, only by studying the value of its dynamical degree \cite{blanccantat}.
This kind of classification can therefore also be of help when checking the integrability of an equation. For example, if an autonomous equation does not have a space of initial conditions and its degree growth is not linear, then one can immediately conclude that the mapping is nonintegrable.

\section*{Acknowledgements}
RW and TM would like to acknowledge support from the Japan Society for the Promotion of Science (JSPS),  through JSPS grants number 18K03355 and 18K13438, respectively.

\end{document}